\documentclass{ifacconf}

\usepackage{xcolor,graphics,graphicx} % for pdf, bitmapped graphics files
\usepackage{epsfig} % for postscript graphics files
\usepackage{natbib}        % required for bibliography
\usepackage{amsmath} % assumes amsmath package installed
\usepackage{amssymb}  % assumes amsmath package installed
\usepackage{amsfonts}
\usepackage{mathtools} 
\usepackage{tikz}
\usepackage{braket}
\usepackage[font=small,skip=0pt]{subcaption}
\usepackage[font=small,skip=0pt]{caption}
\usepackage{tabularray}
\usepackage{url}
\def\ebr#1{{{\color{black}#1}}} % things to review 
\providecommand{\SRG}[1]{\operatorname{SRG}(#1)}

\begin{document}
\begin{frontmatter}

\title{Stability results for MIMO LTI systems via Scaled Relative Graphs\thanksref{footnoteinfo}} 

\thanks[footnoteinfo]{This work was supported by Innovative Tools for Cyber-Physical Energy Systems(InnoCyPES), under the Marie Sklodowska Curie grant 956433.}

\author{Eder~Baron-Prada$^{1,2}$, Alberto Padoan$^{3}$, Adolfo Anta$^{1}$ and~Florian Dörfler$^{2}$} 

\address{$^{1}$Austrian Institute of Technology, Austria. 
\\$^{2}$Automatic Control Laboratory, ETH Zürich, Switzerland.
\\$^{3}$Department of Electrical \& Computer Engineering, University of British Columbia, Canada.}

\begin{abstract}                % Abstract of 50--100 words
This paper proposes a frequency-wise approach for stability analysis of multi-input, multi-output (MIMO) Linear Time-Invariant (LTI) feedback systems through Scaled Relative Graphs (SRGs). Unlike traditional methods, such as the Generalized Nyquist Criterion (GNC), which relies on a coupled analysis that requires the multiplication of models, our approach enables the evaluation of system stability in a decoupled fashion, system by system, each of which is represented by its SRG (or an over-approximation thereof), and it provides an intuitive, visual representation of system behavior. Our results provide conditions for certifying the stability of stable and square MIMO LTI systems connected in closed loop.
\end{abstract}

\begin{keyword}
Input-Output Stability, Linear Systems, Scaled Relative Graphs.
\end{keyword}

\end{frontmatter}
%===============================================================================

\section{Introduction}
In control theory, stability is a fundamental property for the reliable operation of feedback systems across various applications, ranging from industrial automation to power systems \citep{huang2024gain}. The GNC is a traditional method \citep{1102280,skogestad2005} employed as a critical tool in industrial environments. It provides necessary and sufficient conditions for the stability of feedback systems. However, a significant limitation of this classical approach lies in the inherent coupling: Nyquist diagrams require the multiplication of the transfer functions of all systems in the loop to determine stability, which can be difficult in the case of highly-dimensional systems \citep{1102280,zhou1998}.  In addition, in the case of MIMO systems, the challenges posed by coupled analysis become increasingly pronounced \citep{Wang2024}.  The interaction of coupled components complicates stability analysis and problem diagnosis, especially in high-order systems where the Nyquist plot may exhibit multiple loops around the origin, which can signal potential instabilities depending on the system. \ebr{Moreover, the MIMO GNC primarily provides a stability assessment without directly revealing classical gain and phase margins. While the distance of the Nyquist plot from the critical point offers some insight into robustness, establishing a direct correspondence to the scalar margins common in SISO systems remains challenging. This challenge is compounded by the fact that applying the GNC in MIMO settings requires full subsystem models, which are not typically available. } 

We propose a new stability result based on the SRG framework to address these limitations. Originally  introduced in optimization theory for convergence analysis by \cite{ryu2022large}, SRGs have since been applied more broadly, including in the study of stability in nonlinear systems \citep{Chaffey_2023,chaffey2024homotopy}. This approach builds on a homotopy argument, pioneered by ~\cite{Rantzer1997} and later widely adopted for stability analysis in diverse systems~\citep{Rantzer2022,Jonsson2001}.  \ebr{While SRGs for LTI systems were first developed in \cite{pates2021scaled, Chaffey_2023}, our approach leverages a key structural property of LTI operators: the Fourier transform diagonalizes convolution. Consequently, each frequency component evolves independently, enabling a frequency-wise characterization of the SRG. This decomposition, rooted in linearity and time invariance, allows us to replace a global operator-level condition with per-frequency conditions, reducing conservatism relative to global SRG tests and enabling more refined frequency-wise certificates.} \ebr{Independently of this line of work, \cite{chen2025graphicaldominanceanalysislinear} developed graphical tools for assessing dominance of LTI systems. Dominance, however, captures how the system orders nearby trajectories at the differential level, shifting the focus from classical feedback stability to the structure of p-dimensional attracting behaviors. Our contribution instead targets frequency-wise stability assessment and establishes explicit connections with the GNC}. Our result provides a valuable alternative to the traditional GNC, enabling stability analysis in a decoupled manner. Since the SRG is computed using the system's inputs and outputs, it can be estimated directly from measurements, allowing the technique to be applied in a data-driven manner. \ebr{In contrast to the GNC, which requires full subsystem models to construct the loop gain and evaluate its frequency response, the SRG-based approach can operate solely on the systems SRGs or any suitable over-approximation of them, making it more amenable to scenarios involving incomplete models, uncertain dynamics, or purely experimental data.} %This contrasts with traditional methods, such as Lyapunov, eigenvalue analysis, which rely on detailed system models \cite{khalil2002}.

\section{Preliminaries}\label{sec:basics}
The sets of real and complex numbers are denoted by $\mathbb{R}$ and $\mathbb{C}$, respectively. The complex conjugate of $z \in \mathbb{C}$ is denoted by $z^*$. Throughout this work, the real and imaginary parts of a complex number $z \in \mathbb{C}$ are denoted by $\Re(z)$ and $\Im(z)$, respectively. We denote the imaginary unit as $\textup{j}$. % A matrix $ A $ is invertible if there exists a matrix $ A^{-1} $ such that $ AA^{-1} = A^{-1}A = I $, where $ I $ is the identity matrix. 
Let $H$ denote a Hilbert space over the field $F \in \{\mathbb{R},\mathbb{C}\}$. A (bounded) linear operator $A : \mathcal{H} \to \mathcal{H}$ satisfies $A(\alpha x + \beta y) = \alpha A x + \beta A y$ for all $\alpha,\beta \in F$ and $x,y \in \mathcal{H}$. When $\mathcal{H} = \mathbb{C}^n$, we represent such operators by matrices, whose spectrum consists of all $\lambda \in \mathbb{C}$ such that $A - \lambda I$ is not invertible. 

\subsection{Signal Spaces}

We focus on Lebesgue spaces of square-integrable functions $\mathcal{L}_2$.  Given the time axis, $\mathbb{R}_{\ge 0}$, and a field $F \in \{\mathbb{R}, \mathbb{C}\}$, we define the space $\mathcal{L}_2^n(F)$ by the set of signals $u: \mathbb{R}_{\ge 0} \rightarrow F^n$ and $y: \mathbb{R}_{\ge 0} \rightarrow {F}^n$ such that the inner product of $u, y \in \mathcal{L}_2^n({F})$ is defined by $\langle u, y \rangle := \int_0^{\infty} u(t)^* y(t) \, dt$, and the norm of $u$ is defined by $\|u\| := \sqrt{\langle u, u\rangle }<\infty$. The Fourier transform of $u \in \mathcal{L}_2^n({F})$ is defined as $\hat{u}(\textup{j}\omega) := \int_0^{\infty} e^{-\textup{j}\omega t} u(t) \, dt$. \ebr{Finally, the angle between two nonzero vectors $z_1, z_2 \in \mathcal{H}$ is written as $\angle(z_1, z_2):= \arccos\!\left( \tfrac{\Re\langle z_1,\, z_2 \rangle} {\|z_1\|\,\|z_2\|} \right)$}. %Moreover, we define the extension of $\mathcal{L}_2^n({F})$ as $\mathcal{L}_{2,e}^n({F}):=\{u:\mathbb{R}_{\geq0}\rightarrow F^n\;|\;P_Tu\in\mathcal{L}_2\; \forall\; T \in \mathbb{R}_{\geq0}\},$where $P_Tu(t)$ is the truncation operator of the signal $u(t)$ until time $T$. Note that $\mathcal{L}_2^n\subset \mathcal{L}_{2,e}^n$\cite{van_L2_2000}.
 
\subsection{Linear Time-Invariant Systems and Transfer Functions} 
Transfer functions describe the input-output behavior of LTI systems, represented by 
\begin{align*}
\dot{x} = Ax + Bu, \quad y = Cx + Du,
\end{align*}
 where $x\in \mathbb{R}^n$ is the state vector, $u\in \mathbb{R}^m$ is the input, and $y\in \mathbb{R}^p$ is the output, with the matrices $A$, $B$, $C$, and $D$ of appropriate dimension. By applying the Laplace transform with $x(0) = 0$, we derive the system transfer function as $y(s) = H(s)u(s)$\citep{zhou1998}. %An invertible LTI system is defined by a transfer function matrix $ H(s) $ that is non-singular for all $s=\textup{j}\omega$, with $\omega\in \mathbb{R}$, i.e., $ \det(H(\textup{j}\omega)) \neq 0 $, ensuring the existence of a unique inverse transfer function $ H^{-1}(\textup{j}\omega) $ satisfying $ H(\textup{j}\omega)H^{-1}(\textup{j}\omega) = H^{-1}(\textup{j}\omega)H(\textup{j}\omega) = I_n $\cite{skogestad2005}. 
 This work focuses on the space $ \mathcal{RH}_\infty $ of rational, proper, and stable transfer functions, which describe bounded, causal LTI systems. These systems define an input-output gain that measures the ratio of the output size to the input. For $ \mathcal{L}_{2} $ signals, this gain is equivalent to the $ H_\infty $ norm \citep{Chen2022, zhou1998}.  Moreover, finite incremental gain and asymptotic stability of all input/output trajectories are equivalent for an operator derived from a dynamical system, provided the system is reachable and observable\citep{Fromion_1996}.% Finally, the relationship between $\mathcal{L}_{2}$-stability and finite incremental gain does not hold for nonlinear systems where superposition fails.

\subsection{Generalized Nyquist Criterion (GNC)}
The GNC is a fundamental stability method used in control theory to determine if a closed-loop system is stable based on the open-loop frequency response. % {An adapted version is presented here since we are considering specific systems in the space $\mathcal{RH}_\infty$.}
\begin{figure}[hbt]
    \centering
    \begin{tikzpicture}[scale=1, every node/.style={transform shape}]
% The feedback system
\draw (9.75,4.3) rectangle (11.25,3.7);
\node at (10.5,4) {$H_1$};
\draw (9.75,3.6) rectangle (11.25,3);
\node at (10.5,3.3) {$H_2$};
\draw[-latex, line width = .5 pt] (12.5,4) -- (13.5,4);
\draw[-latex, line width = .5 pt] (11.25,4) -- (12.5,4) -- (12.5,3.3) -- (11.25,3.3);
\draw[fill = white] (8.5,4) circle [radius=0.05];
\draw[-latex, line width = .5 pt] (9.75,3.3) -- (8.5,3.3) -- (8.5,3.95) ;
\draw[-latex, line width = .5 pt] (8.55,4) -- (9.75,4);
\draw[-latex, line width = .5 pt] (7.5,4) -- (8.45,4);
%% letters
\node at (8.6892,3.8444) {\tiny$-$};
\node at (8.3332,4.1664) {\tiny$+$};
\node at (9,4.1664) {$e$};
\node at (8,4.1664) {$u$};
\node at (13,4.1664) {$y$};
\end{tikzpicture}
\caption{Feedback interconnection between $H_1$ and $H_2$. }
    \label{fig:fb}
\end{figure}

\begin{theorem}\textbf{(GNC)\citep{1102280}}\label{thm:GNC} Consider the feedback interconnection in Fig. \ref{fig:fb}. Assume $H_1(s),H_2(s) \in \mathcal{RH}_\infty^{m\times m}$ 
 and the system interconnection is well-posed. The closed-loop system is exponentially stable if and only if
\begin{align}
    \operatorname{det}(I+H_1(s)H_2(s))\neq0 , \forall  \omega \in \mathbb{R}\label{eqn:GNC}
\end{align}
and the winding number of $(I+H_1(s)H_2(s))$ around the origin is zero.
\end{theorem}

\ebr{ In Theorem \ref{thm:GNC} and throughout this paper, we use the concept of well-posedness as defined in \citep[Section.~5.2]{zhou1998}, and the winding number is computed following the approach outlined in \citep[Lemma~4.8]{skogestad2005}}. The winding number is an integer indicating the net number of times a closed curve encircles a point, counting counterclockwise encirclements as +1 and clockwise encirclements as -1. It is computed as the total signed rotations around the point. The GNC comprises two key conditions: the first is a frequency-wise condition, as expressed in \eqref{eqn:GNC}, while the second involves evaluating the trajectory of \eqref{eqn:GNC} over the entire frequency range\ebr{ , i.e., counting encirclements}. A simplified sufficient version of the GNC could be formulated as follows.

\begin{theorem}\textbf{(Sufficient GNC)\citep{GRIGGS_2012_sufficientNyquist}}\label{thm:sufficient_GNC} Consider the feedback interconnection in Fig. \ref{fig:fb}. Assume $H_1(s),H_2(s) \in \mathcal{RH}_\infty^{m\times m}$, and the system interconnection is well-posed. If 
\begin{align}
    \operatorname{det}(I+\tau H_1(s)H_2(s))\neq0, \forall \tau \in (0,1],\forall \omega \in \mathbb{R},\label{eqn:sufficient_GNC}
\end{align}
 then the closed-loop system is exponentially stable.
\end{theorem}

Theorem \ref{thm:sufficient_GNC} provides a sufficient condition for exponential stability. Broadly speaking, it implies that if $\operatorname{det}(I + H_1(s)H_2(s))$ intersects the negative real axis, an encirclement occurs \citep{GRIGGS_2012_sufficientNyquist}. However, since the theorem does not explicitly track the evolution of the determinant trajectory across the frequency spectrum, it cannot determine if the winding number of the trajectory around the origin is zero. In scenarios where the models of $H_1(s)$ and $H_2(s)$ are unknown, Theorem \ref{thm:sufficient_GNC} serves as the key tool to ensure closed-loop stability. However, similar to classical GNC, Theorem \ref{thm:sufficient_GNC} requires the multiplication of the transfer functions for both systems — a condition that can be limiting and prone to numerical inaccuracies. To address this limitation, we instead leverage SRGs, which enable decoupled stability conditions, as demonstrated in the following section.

\section{Scaled Relative Graphs}\label{sec:SRG}
SRGs were first introduced by \cite{ryu2022large} as an analytical tool in optimization theory. We recall its definition and specialize it for linear systems. \ebr{ Let $A : \mathcal{H} \rightarrow \mathcal{H}$ be an operator. In analogy with the definition of the graph of a causal system, we define the \emph{graph} of $A$ by
\[
\mathcal{G}(A) := 
\left\{
\begin{bmatrix}
u \\[2mm]
A u
\end{bmatrix}
\in \mathcal{H} \times \mathcal{H}
\right\}.
\]
For any pair of inputs $u_{1}, u_{2} \in \mathcal{H}$, their corresponding outputs are 
$y_{1} = A(u_{1})$ and $y_{2} = A(u_{2})$, and thus
\[
\begin{bmatrix} u_{1} \\ y_{1} \end{bmatrix},
\,
\begin{bmatrix} u_{2} \\ y_{2} \end{bmatrix}
\in \mathcal{G}(A).\]}
\ebr{Then, the $\SRG{A}$ is the set of all complex numbers obtained from every pair of distinct inputs $u_1, \; u_2\in \mathcal{H}$, defined as}
\begin{align}\label{eq:SRG_general}
\operatorname{SRG}(A)
:= 
\Big\{ 
\tfrac{\|y_2 - y_1\|}{\|u_2 - u_1\|}
\, e^{\pm j \theta}:&\\
\nonumber\theta = \angle&( y_2 - y_1, u_2 - u_1)\Big\}.
\end{align} 
\ebr{ If $A : \mathbb{C}^n \to \mathbb{C}^n$ is linear, then for all $u \in \mathbb{C}^n \setminus \{0\}$. The incremental formulation in \eqref{eq:SRG_general} is equivalent to }
\begin{align}
\operatorname{SRG}(A)=\Big\{
\tfrac{\|y\|}{\|u\|}
\, e^{\pm j \theta}: 
\theta = \angle (y,u)\Big\}
,
\label{eq:SRG_linear}
\end{align}
\ebr{which follows by linearity, setting $u_1 = 0$ and $u_2 = u$ in \eqref{eq:SRG_general}. In case there exists a $u\neq0$ such that  $y=0$, then $\SRG{A}$ contains the 0 element. The ratio $\frac{\|y\|}{\|u\|}$ represents the amplitude change in the output compared to the input. The term $\angle (y,u)$ indicates the angle between the input and the output \citep{millman1993}. }
We focus exclusively on square transfer functions, meaning the systems under consideration have equal inputs and outputs ($m=p$), since angular differences between vectors are  {classically} defined for vectors of the same dimension\citep{millman1993}. Such systems are typical in power systems \citep{huang2024gain} and aircraft/drone systems \citep{Marani2022}, among others. Note that $\operatorname{SRG}(A)$ is symmetric with respect to the real axis.% More details about the relation between \eqref{eqn:SRG_operator_nonlinear} and \eqref{eqn:SRG_operator} can be found in Appendix \ref{appendix:L2properties}.

\subsection{SRG properties of operators in Hilbert Spaces } \label{subsec:properties}
We recall three SRG properties: the SRG inverse, the operator class, and the chord property,  \ebr{ adapted from \cite{ryu2022large}}. We define inversion in the complex plane by the Möbius transformation
\begin{fact}[Inversion of a SRG]
 If $A$ is an operator, then $\operatorname{SRG}(A^{-1})=\operatorname{SRG}(A)^{-1}=\{(z^{-1})^*\mid z\in \operatorname{SRG}(A)\}$. \ebr{In particular, when $0 \in \operatorname{SRG}(A)$, we define its inverse to be $\infty$.} 
\end{fact}
\begin{fact}[Operator Class]
 Given an operators class $\mathcal{A}$, the SRG of $\mathcal{A}$ is given by $ \operatorname{SRG}(\mathcal{A}) := \bigcup_{A \in \mathcal{A}} \operatorname{SRG}(A)$.
\end{fact}
Note that a class does not need any structure. In addition, a class can consist of a single operator.
\begin{fact}[Chord Property]
An operator $A$ is said to satisfy the chord property if for every bounded $ z \in \operatorname{SRG}(A)$, the line segment $[z, z^*]$, defined as $z_1,z_2 \in \mathbb{C}$ as $[z_1,z_2] := \{ \beta z_1 + (1 - \beta)z_2 \mid \beta \in [0,1] \}$, is in $\operatorname{SRG}(A)$. 
\end{fact}
We denote by $\overline{A}$ an operator satisfying the chord property such that $\operatorname{SRG}(A) \subseteq \operatorname{SRG}(\overline{A})$. Generally, the SRG does not inherently satisfy the chord property. Note that the over-approximation is not unique, but we could obtain a tight approximation as

\begin{definition}[Tight chord approximation] 
\label{def:approx_chord}
    \ebr{Given $A$, $\overline{A}$ is defined as an operator such that  $ \operatorname{SRG}(\overline{A})\supseteq\operatorname{SRG}(A)$, and}
\begin{align*}
     \operatorname{SRG}(\overline{A}):=\Set{[z,z^*]\;\forall\;z\in\SRG{A} }.
\end{align*}
\end{definition}
\ebr{Note that several other approximations can be made. In fact, in \cite{Baron2025_decentralized} and in \cite{de2025dissipativity} circles are used to over-approximate the SRG of an operator, which also meet the chord property.}

\subsection{Generalized Feedback Stability Theorem (GFT)}
We now recall the GFT in the Theorem \ref{thm:GFT} originally proposed in \cite{Chaffey_2023} and later extended in \cite{chaffey2024homotopy}.
\begin{theorem}\textbf{(GFT)}\label{thm:GFT} Consider the feedback interconnection in Fig. \ref{fig:fb} between any pair of operators $A_1 \in \mathcal{A}_1$ and $A_2 \in \mathcal{A}_2$, where $\mathcal{A}_1,\mathcal{A}_2 \in \mathcal{L}_{2}$ are a class of operators with finite incremental gain. If, for all $\tau \in (0, 1]$,
\begin{align}
     \operatorname{SRG}(\mathcal{A}_1)^{-1} \cap -\tau \operatorname{SRG}(\overline{\mathcal{A}}_2) = \emptyset,
     \label{eqn:SRGEqui}
\end{align}
then the feedback interconnection map is $\mathcal{L}_{2}$ stable.
\end{theorem}

 The selection of which SRG to approximate to satisfy the chord property is arbitrary, as discussed in \cite{Chaffey_2023}. If finite incremental gain is required,  one needs strict separation between the SRGs \citep{chaffey2024homotopy}. \ebr{Notably, finite incremental gain is guaranteed if \eqref{eqn:SRGEqui} is met for LTI systems}. The strict separation between $ \SRG{\mathcal{A}_1} $ and $ \SRG{\mathcal{A}_2} $ is said to hold if the infimum of the absolute distances between any point $ a_1 \in \SRG{\mathcal{A}_1} $ and any point $ a_2 \in -\SRG{\overline{\mathcal{A}}_2} $, satisfies
\begin{align}
    \inf \left\{ |a_1 - a_2| \mid a_1 \in \SRG{\mathcal{A}_1}, \, a_2 \in -\SRG{\overline{\mathcal{A}}_2} \right\} > 0. \label{eqn:margin}
\end{align}
% one can assume $a_1 \in \SRG{\mathcal{A}_1}$ and ${a_2 \in -\SRG{\overline{\mathcal{A}}_2}}$, with the distance between the SRGs defined as $\inf|a_1 - a_2|$, which must be positive\cite{chaffey2024homotopy}.

%%%%%%%%%%%%%%%%%%%%%%%%%%%%%%%%%%%%%%%
\section{Stability Conditions based on SRGs }\label{sec:SRGconditions} 
In this section,  {we introduce the frequency-wise version of the SRG for LTI systems and Theorem \ref{thm:GFTLTI} for evaluating the stability of feedback systems}.  {Finally, we provide conditions under which the Theorem \ref{thm:sufficient_GNC} and Theorem \ref{thm:GFTLTI} are equivalent}. 
 
\subsection{SRG of LTI systems}
\ebr{A transfer function $H(s)$ can be viewed as a collection of frequency-specific operators, each describing the system’s response at a particular frequency $\omega \in \mathbb{R}$ \citep{zhou1998}. Because convolution is diagonalized by the Fourier transform, the evolution of each frequency component is independent. Consequently, the action of an LTI operator decouples across frequencies, enabling a frequency-wise characterization of the SRG. This allows stability to be assessed by examining the system’s response at each individual frequency rather than analyzing the full transfer function as a whole.} The SRG allows us to calculate a set of gains and phases for all possible inputs for each operator. By plotting all individual SRGs in a 3D space, we capture the full range of dynamic behaviors of $H(s)$.

The shape of the SRG for LTI systems varies depending on the system dimension. For SISO systems, the SRG at each frequency consists of two symmetrical points about the real axis \cite{pates2021scaled}. \ebr{In MIMO systems, the SRG forms either two symmetrical curves or two points when the system input $ m $ is two.} For $ m \geq 3 $, the SRG can expand into two symmetrical closed regions in the complex plane \citep{pates2021scaled}. Generally, the SRG consists of two disconnected, symmetrical sets across the real axis. However, these sets become connected only if the SRG contains points on the real axis, which occurs when the system has real eigenvalues\cite[Proposition 5]{hannah2016scaled}.
%  \begin{remark}[Chord Property of LTI systems]
% An LTI system has the chord property if for each $\omega\in [0,\infty)$ and for every $ z \in \operatorname{SRG}(H(s))$, the line segment $[z, z^*]$, defined as $z_1,z_2 \in \mathbb{C}$ as $[z_1,z_2] := \{ \beta z_1 + (1 - \beta)z_2 \mid \beta \in [0,1] \}$, is in $\operatorname{SRG}(H(s))$ .
%  \end{remark}

\subsection{Decoupled Stability Theorem on LTI systems}
This subsection uses the GFT to derive stability conditions based on SRGs for LTI systems. We present a frequency-wise interpretation of Theorem~\ref{thm:GFT} tailored specifically for LTI systems as follows

\begin{theorem}\label{thm:GFTLTI} (\textbf{GFT for LTI systems})
Consider the feedback interconnection in Fig. \ref{fig:fb}. Assume $H_1(s),H_2(s) \in \mathcal{RH}_\infty^{m\times m}$, and  either $H_1(s)$ or $H_2(s)$ satisfies the {chord property}.  If, $\; \forall \omega \in [0, \infty)$,
\begin{align}
     \operatorname{SRG}({H}_1(s))^{-1} \cap -\tau\operatorname{SRG}({{H}}_2(s)) = \emptyset,\forall \tau \in (0,1],
     \label{eqn:SRGEquiLTI}
\end{align}
then the closed-loop system is $\mathcal{L}_{2}$ stable.
\end{theorem}

\begin{pf}
      {The proof can be found in the Appendix} \ref{proof:GFTLTI} .
\end{pf} 

 Theorem \ref{thm:GFTLTI} shows that system stability can be evaluated by comparing the SRGs of each system at every frequency $\omega\in[0, \infty)$. \ebr{ Theorem \ref{thm:GFTLTI} strengthens Theorem 4 in \cite{Chaffey_2023}, which defines an operator that is calculated considering the hyperbolic convex hull of $\SRG{H_1(s)} \forall \omega \in [0,\infty)$ for all frequencies simultaneously, i.e., Theorem 4 in \cite{Chaffey_2023} require handling interactions across different frequencies, whereas Theorem \ref{thm:GFTLTI} removes this constraint. As we are using a homotopy condition on linear systems, well-posedness follows \cite[Thm.~7.40]{scherer2000linear}.} By allowing the system response to be represented as a collection of independent operators at each frequency for both SISO and MIMO systems, Theorem \ref{thm:GFTLTI} simplifies the stability analysis. \ebr{ A related condition has been developed independently by \cite{chen2025graphicaldominanceanalysislinear}, where the focus is on the calculation of the resulting dominance of the system.}  Additionally, the coefficient $\tau$ can be multiplied to either system.

\subsection{Computational methods for SRG separation}

\ebr{Once the SRGs of $H_1(s)$ and $H_2(s)$ are obtained at each frequency, stability reduces to checking whether the two sets remain disjoint. This can be done using computational methods of different accuracy, including:}

\subsubsection{Disk approximation:}
\ebr{The simplest approach approximates each SRG by a disk $\mathbb{D}(c_k,r_k)$ characterized by its center $c_k$ and radius $r_k$. These parameters can be obtained from the sampled SRG boundary at $N_\phi$ angular directions as the mean and maximal radial deviation, with cost $\mathcal{O}(N_\phi)$ per frequency. Once computed, separation is verified through $|c_1-c_2| > r_1 + r_2$, an $\mathcal{O}(1)$ test. This yields the fastest but most conservative assessment.}

\subsubsection{Convex-hull distance:}
\ebr{A more accurate check samples each SRG boundary and constructs convex hulls $S_k=\operatorname{conv}(\SRG{H_k(s)})$. The  distance can be computed using the \emph{Gilbert--Johnson--Keerthi (GJK)} algorithm or equivalent convex-distance methods. The per-frequency complexity is $\mathcal{O}(N_\phi\log N_\phi)$ for hull construction, and the complexity for distance evaluation is $\mathcal{O}(N_{\phi})$.}

\subsubsection{Naïve pointwise evaluation:}
\ebr{As a reference, an exhaustive distance evaluation computes all pairwise distances between all angular points of $\operatorname{SRG}(H_1(s))^{-1}$ and $\operatorname{SRG}(H_2(s))$. The per-frequency complexity is $\mathcal{O}( N_\phi^2)$. This brute-force strategy provides the exact distance but is computationally prohibitive for fine discretizations.}

\ebr{All previous computational approaches include off-the-shelf routines that can be implemented in several programming environments. Consequently, the total complexity across the considered frequency range grows only linearly with the number of frequency samples $N_\omega$.}

\subsection{Equivalence between GFT and sufficient GNC }

Our main result establishes conditions for  $H_1$ and $H_2$ under which Theorem \ref{thm:sufficient_GNC} and Theorem \ref{thm:GFTLTI} are equivalent, ensuring $\mathcal{L}_{2}$ stability.  {This analysis focuses exclusively on the interconnection of stable systems, explicitly excluding unstable systems from consideration.}

\begin{theorem}(\textbf{Equivalence between GFT and GNC})\label{thm:equivalence}
Consider the feedback interconnection in Fig. \ref{fig:fb}. Assume $H_1(s),H_2(s) \in \mathcal{RH}_\infty^{m\times m}$  and the system interconnection is well-posed, and either $H_1(s)$ or $H_2(s)$ has the {chord property}, \ebr{ and $H_1(s)$ is invertible}. Then, the following statements are equivalent:
\begin{enumerate}
    \item $\operatorname{det}(I+\tau H_1(s)H_2(s))\neq0 \;\forall \;\tau \in (0,1]$,  $\forall \omega \in \mathbb{R}$.
    \item $\operatorname{SRG}({H}_1(s))^{-1} \cap -\tau \operatorname{SRG}({{H}}_2(s)) = \emptyset$, $\; \forall\tau \in (0,1]$,  $\forall \omega \in [0,\infty)$.
\end{enumerate}
\end{theorem}
\begin{pf}
     {The proof can be found in the Appendix} \ref{proof:equivalence}.
\end{pf}

Theorem \ref{thm:equivalence} provides a frequency-wise sufficient condition for verifying the stability of MIMO LTI systems. By bridging Theorem~\ref{thm:sufficient_GNC} and Theorem \ref{thm:GFTLTI}, it establishes their equivalence in guaranteeing $\mathcal{L}_2$-stability.  Importantly, it demonstrates that SRGs can serve as an alternative to the GNC for stability analysis, offering a complementary approach for examining the stability of feedback systems. 
\subsection{Comparison Between SRG-Based and Nyquist Stability Criteria}

\ebr{The sufficient GNC remains a cornerstone of frequency-domain stability analysis, and when a full rational model of the loop gain is available it provides an exact algebraic certification. In practical settings, however, such complete models are rarely known: engineers typically rely on sampled frequency-response data rather than closed-form transfer matrices. Under these circumstances the Nyquist route becomes less direct, since one must either reconstruct a model to apply the algebraic tests, or approximate the Nyquist contour from data, a process that can obscure subsystem-level contributions and still requires evaluating $\det(I+L(s))$ across the frequency grid. For MIMO systems in particular, this demands $\mathcal{O}(N_\omega n^3)$ operations and merges all subsystem dynamics into a single loop-gain object whose encirclement plot becomes increasingly difficult to interpret as dimension and uncertainty grow.}

\ebr{The SRG framework, by contrast, is naturally suited to situations where only frequency samples are available. Rather than collapsing the subsystems into a single loop gain, the SRG test checks a frequency-wise geometric separation between $\operatorname{SRG}(H_1(s))^{-1}$ and $-\tau,\operatorname{SRG}(H_2(s))$. This preserves the individual gain–phase structure of each component and enables stability certification directly from data (no identification step or full model reconstruction is required). Even coarse over-approximations of SRGs can suffice, making the method particularly attractive when subsystem models are incomplete or uncertain. Although computing SRG boundaries scales as $\mathcal{O}(N_\omega N_\phi n^3)$, the resulting conditions remain interpretable and modular, revealing stability margins that are often obscured in MIMO Nyquist plots.}

\ebr{Overall, the distinction between the two approaches becomes most pronounced in data-driven settings. Algebraic Nyquist criteria excel when a full model is in hand, but absent such a model they provide no direct route from raw frequency samples to a stability verdict. SRGs, on the other hand, are designed to operate precisely in this regime: they enable the evaluation of stability conditions from sampled responses without reconstructing the underlying dynamics. This makes SRG-based tests not merely an alternative to Nyquist reasoning, but a more natural tool when stability must be certified from experimentally obtained frequency data.}

\section{Numerical Examples}\label{sec:examples}

\subsection{Stability of a MIMO system with $m=2$}
 {In this section, we aim to prove the efficiency of Theorem \ref{thm:GFTLTI} and Theorem \ref{thm:sufficient_GNC}.} Consider $H_1(s)$ and $H_{2}(s)$ as
\begin{align}
H_1(s)&=\left[\begin{array}{cc} \frac{50\,s+2500}{s^2+100\,s+2501} & \frac{50}{s^2+100\,s+2501}\\ \frac{30}{s^2+100\,s+2501} & \frac{30\,s+2501}{s^2+100\,s+2501} \end{array}\right],\label{eqn:H2}\\
H_{2}(s)&=\left[\begin{array}{cc}
\frac{2s+1}{(s+10)^3} & \frac{s+12}{(s+1)^2}\\
\frac{5s+10}{(s+15)^3} & \frac{s+22}{(s+6)(s+10)^2}
\end{array}\right]\label{eqn:H6}.
\end{align}

\begin{figure}[htb]
    \centering
\begin{subfigure}[b]{0.24\textwidth}
    \centering
    \includegraphics[width=1\linewidth]{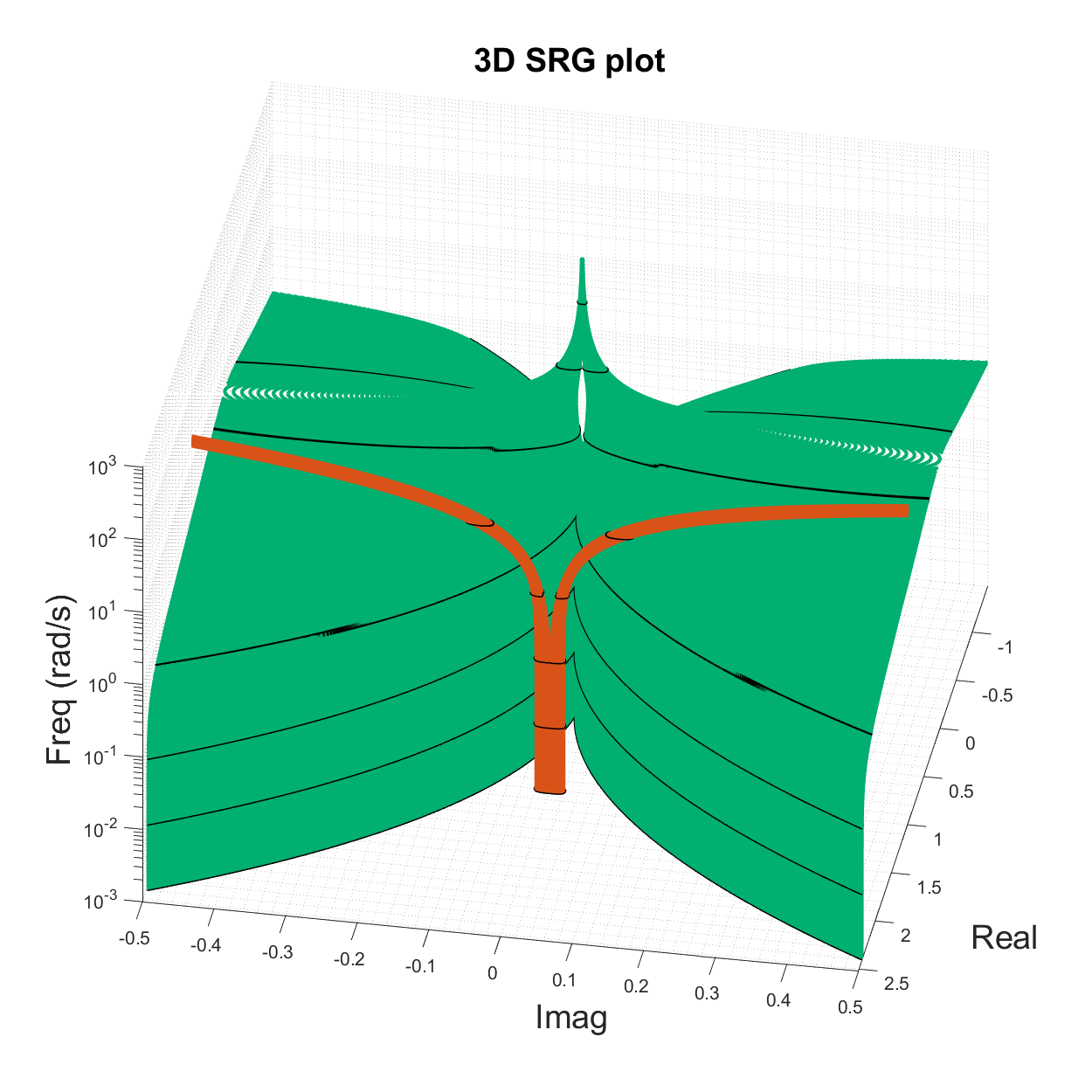}
    \caption{} \label{fig:ex3_SRG3D}
\end{subfigure}  
\begin{subfigure}[b]{0.24\textwidth}
    \centering
    \includegraphics[width=1\linewidth]{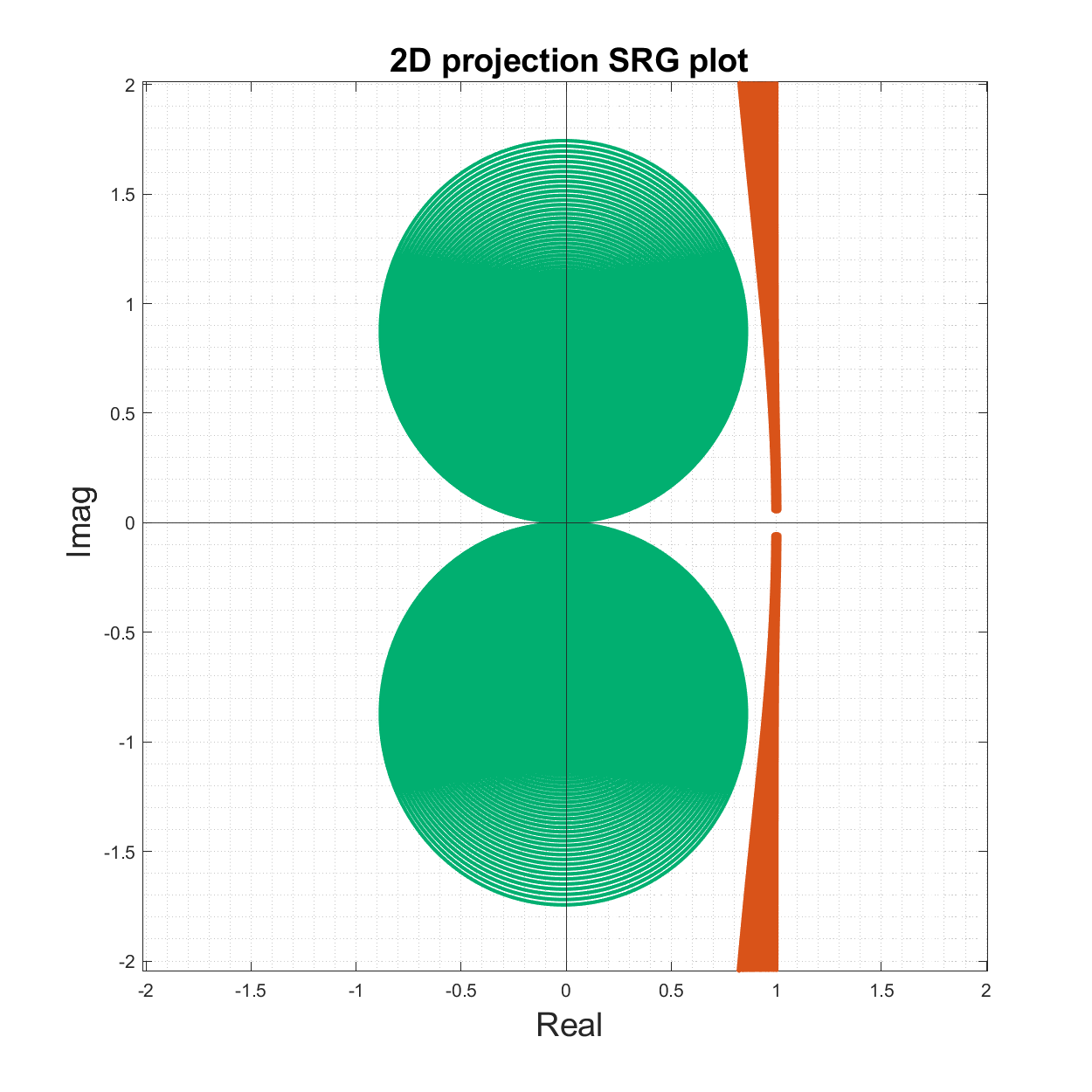}
    \caption{} \label{fig:ex3_2D_a}
\end{subfigure}   
\begin{subfigure}[b]{0.24\textwidth}
    \centering
    \includegraphics[width=1\linewidth]{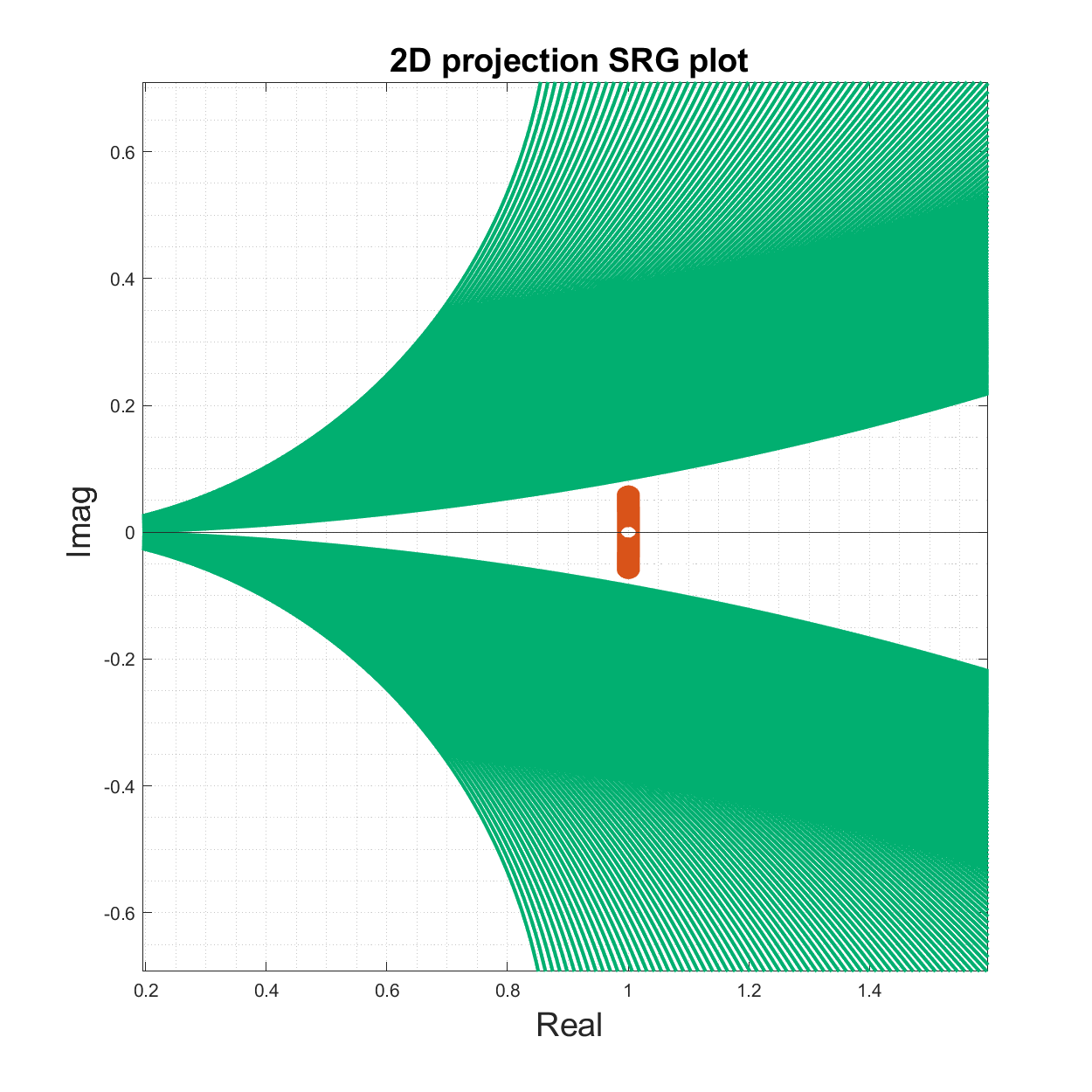}
    \caption{} \label{fig:ex3_2D_b}
\end{subfigure}
\begin{subfigure}[b]{0.24\textwidth}
    \centering
    \includegraphics[width=1\linewidth]{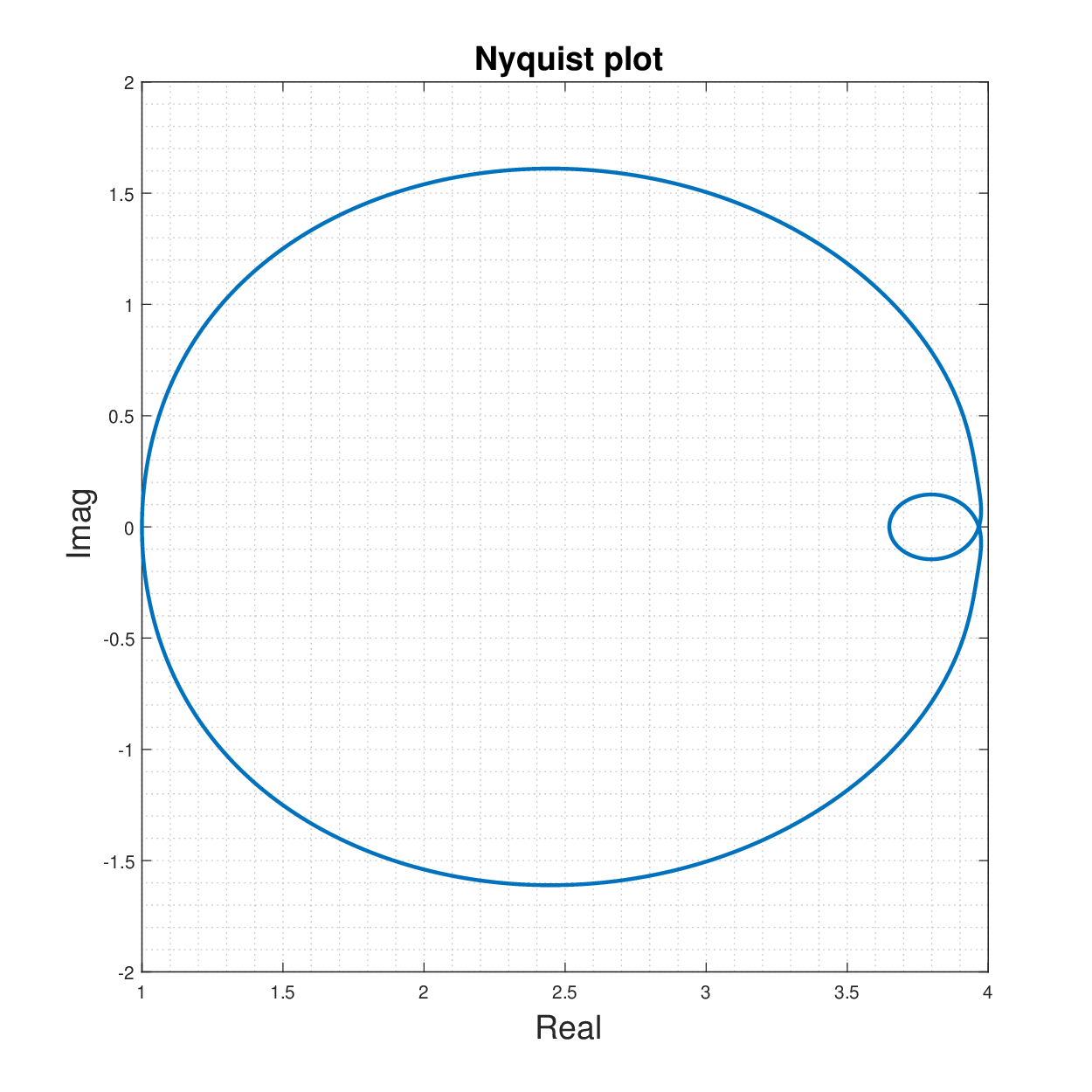}
    \caption{} \label{fig:ex3_Nyquist}
\end{subfigure} 
    \caption{$-\operatorname{SRG}(H_{2}(s))$  in green, $\operatorname{SRG}(H_1^{-1}(s))$ in orange $~\forall \omega\in[10^{-3},10^3]$rad/s. (a) 3D SRGs visualization. (b)~2D projection of the SRGs in the complex plane for $\omega\in[10^{-3},1]$~rad/s. (c)~2D projection of the SRGs in the complex plane for $\omega\in[1,10^{3}]$~rad/s.  (d) Nyquist plot using $\det(I+H_{1}(s)H_2(s))$.}
    \label{fig:SRGExamples3}
\end{figure}

Fig. \ref{fig:SRGExamples3} presents $-\tau\SRG{H_2(s)} $ with $\tau=1$ and $\SRG{ H_1^{-1}(s)} $. Fig. \ref{fig:ex3_SRG3D} provides a 3D plot of the SRG for both operators, though this plot alone does not yield conclusive insights. To facilitate clearer interpretation, we include two projections onto the complex plane for two different frequency ranges in Fig.~\ref{fig:ex3_2D_a} and Fig.~\ref{fig:ex3_2D_b}. { As shown in Figures \ref{fig:ex3_SRG3D}, \ref{fig:ex3_2D_a}, and \ref{fig:ex3_2D_b}, $ -\SRG{H_2(s)} $ exhibits varying area in the complex plane across the frequency spectrum, expanding to a larger region at low frequencies ($ \omega < 1 $ rad/s) and contracting at higher frequencies ($ \omega > 1 $ rad/s).}

Conversely, $ \operatorname{SRG}(H_1^{-1}(s)) $ exhibits an opposite pattern to $ -\operatorname{SRG}(H_{2}(s)) $. Specifically, Figure \ref{fig:ex3_2D_a} shows the low-frequency behavior ($ \omega < 1 $ rad/s), where $ \operatorname{SRG}(H_2(s)) $ spans a smaller region compared with its behavior at higher frequencies ($ \omega > 1 $ rad/s). Furthermore, Theorem \ref{thm:GFTLTI}  shows that the system is stable, given that $ -\tau\operatorname{SRG}(H_{2}(s)) \forall \tau\in(0,1]$ does not intersect $ \operatorname{SRG}(H_1^{-1}(s)) $ at any frequency. %The Nyquist plot shown in Fig.\ref{fig:ex3_Nyquist} can also guarantee the $\mathcal{L}_2$ stability. %as the SRG plots. In this case, as we know there are no encirclements to the origin, Theorem \ref{thm:equivalence_full} could be applied. Moreover, the decoupled analysis of SRGs allows for a direct comparison of different systems in potential feedback configurations, meaning we can evaluate one system against others that might be introduced in a feedback loop. Furthermore, SRG analysis identifies specific frequency ranges that may benefit from targeted adjustments, providing clearer guidance on where to reinforce system stability.

\subsection{Stability of a high order MIMO system with $m=3$} \label{example:MIMOn3}
Consider a square system, $H(s)$,  with $m=3$ of order $50$, i.e.,  with $50$ stable poles, generated by the Matlab command \texttt{rss(50,3,3)}, with no unstable zeros\footnote{The system under analysis can be found at \url{https://github.com/eder-baron/SRG-ECC-2025}}. Furthermore, consider $H_{3}(s)$ and $H_{4}(s)$ as \eqref{eqn:H4} and \eqref{eqn:H5}.
\begin{footnotesize}
\begin{align}
    H_{3}(s)&=\left[\begin{array}{ccc} \frac{33\,\left(s+1\right)}{{\left(s+\frac{143}{10}\right)}^2} & \frac{18\,\left(s+14\right)}{5\,{\left(s+15\right)}^2} & \frac{21\,\left(s+\frac{23}{10}\right)}{5\,{\left(s+15\right)}^2}\\ \frac{36\,\left(s+2\right)}{\left(s+14\right)\,\left(s+55\right)} & \frac{39\,\left(s+13\right)}{\left(s+15\right)\,\left(s+\frac{27}{2}\right)} & \frac{30\,\left(s+2\right)}{{\left(s+15\right)}^2}\\ \frac{30\,\left(s+\frac{3}{2}\right)}{{\left(s+7\right)}^2} & \frac{18\,\left(s+\frac{5}{2}\right)}{5\,\left(s+4\right)\,\left(s+\frac{7}{2}\right)} & \frac{39\,\left(s+3\right)}{{\left(s+15\right)}^2} \end{array}\right]
\label{eqn:H4},\\
    H_{4}(s)&=-\left[\begin{array}{ccc} \frac{88\,\left(s+1\right)}{{\left(s+\frac{143}{10}\right)}^2} & \frac{48\,\left(s+14\right)}{5\,{\left(s+15\right)}^2} & \frac{56\,\left(s+\frac{23}{10}\right)}{5\,{\left(s+15\right)}^2}\\ \frac{96\,\left(s+2\right)}{\left(s+14\right)\,\left(s+55\right)} & \frac{104\,\left(s+13\right)}{\left(s+15\right)\,\left(s+\frac{27}{2}\right)} & \frac{80\,\left(s+2\right)}{{\left(s+15\right)}^2}\\ \frac{80\,\left(s+\frac{3}{2}\right)}{{\left(s+7\right)}^2} & \frac{48\,\left(s+\frac{5}{2}\right)}{5\,\left(s+24\right)\,\left(s+\frac{27}{2}\right)} & \frac{104\,\left(s+3\right)}{{\left(s+15\right)}^2} \end{array}\right]
\label{eqn:H5}.
\end{align}
\end{footnotesize}
% 
%We aim to demonstrate the benefits of using SRG-based conditions for decoupled system analysis, emphasizing the advantages of decoupling characteristics. This approach contrasts with the Nyquist criterion, which can become complex and challenging to interpret when multiple clockwise and counterclockwise loops are present. 

%
\begin{figure}[htb]
    \centering 
\begin{subfigure}[b]{0.24\textwidth}
    \centering    \includegraphics[width=1\linewidth]{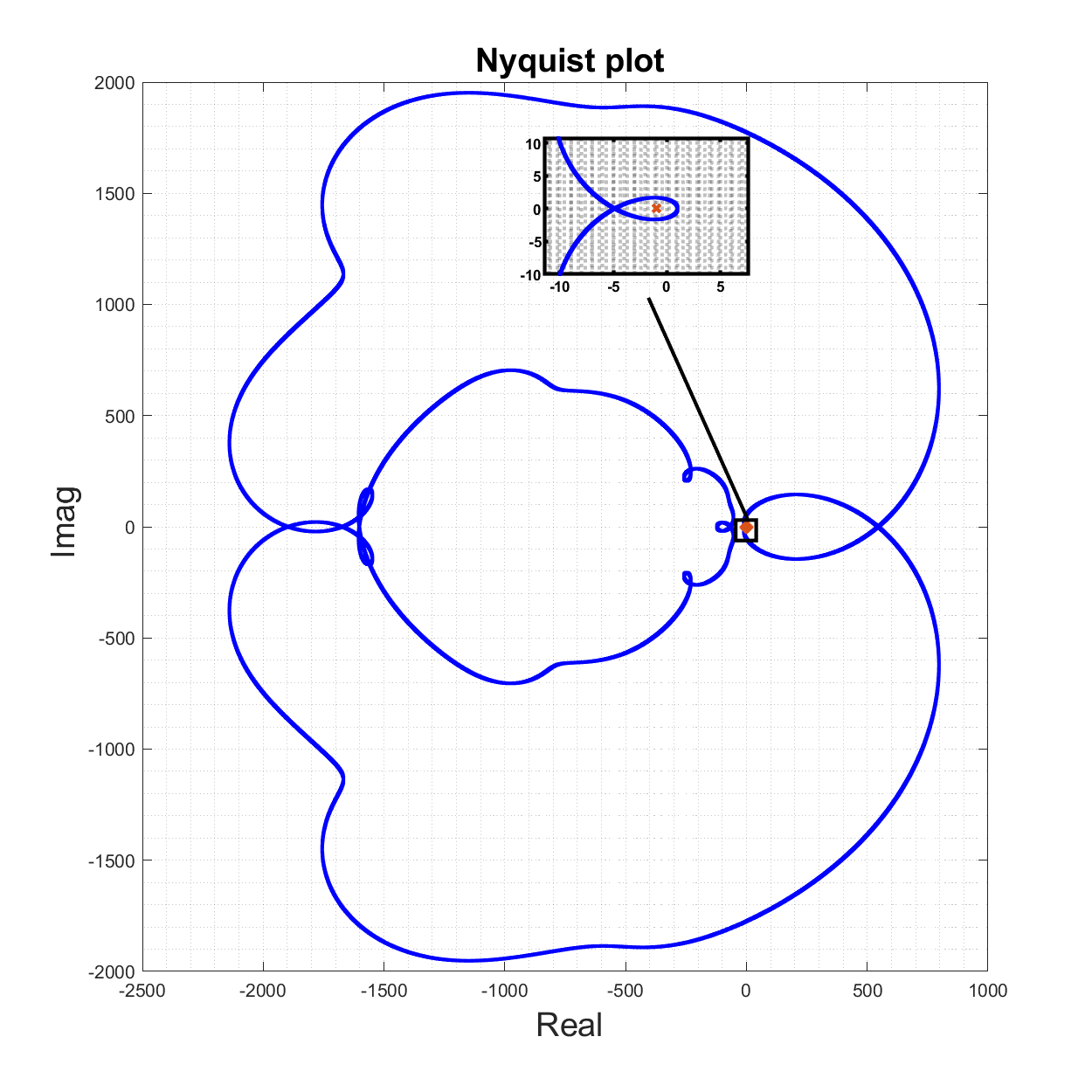}
    \caption{} \label{fig:ex2_uns_nyquist}
\end{subfigure} 
\begin{subfigure}[b]{0.24\textwidth}
    \centering 
    \includegraphics[width=1\linewidth]{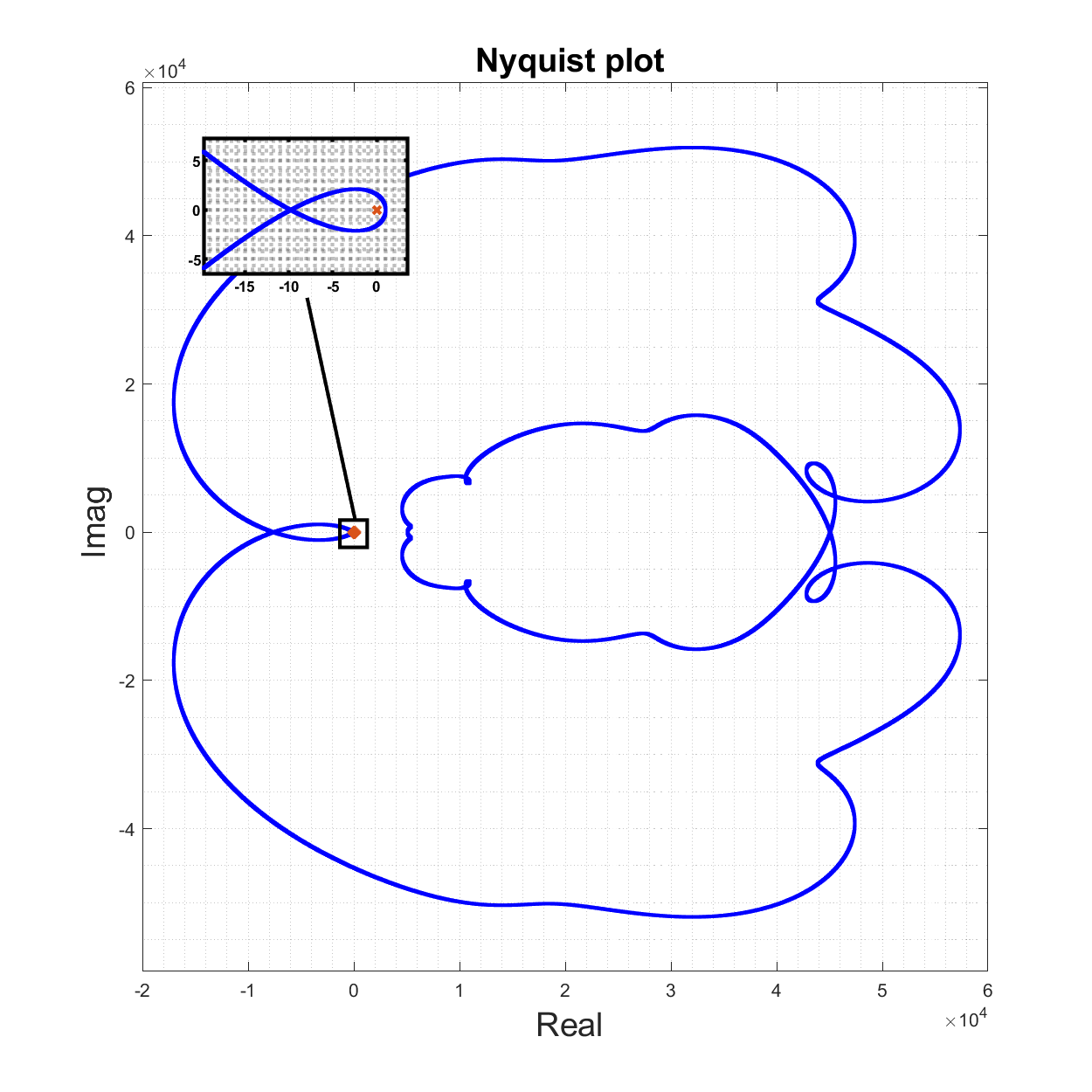}
    \caption{} \label{fig:ex2_st_nyquist}
\end{subfigure} 
\begin{subfigure}[b]{0.24\textwidth}
    \centering
    \includegraphics[width=1\linewidth]{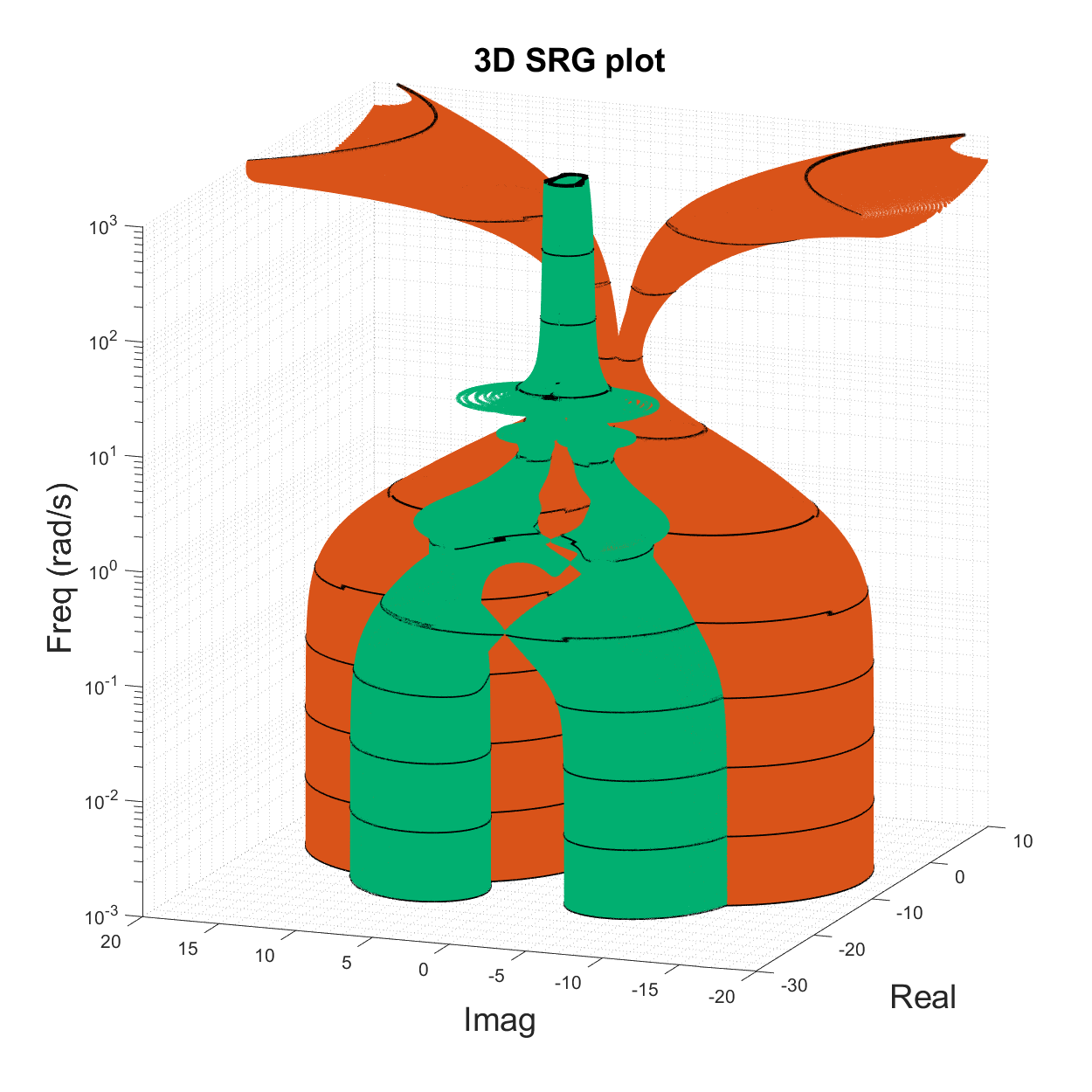}
    \caption{} \label{fig:ex2_uns_3D}
\end{subfigure} 
\begin{subfigure}[b]{0.24\textwidth}
    \centering
    \includegraphics[width=1    \linewidth]{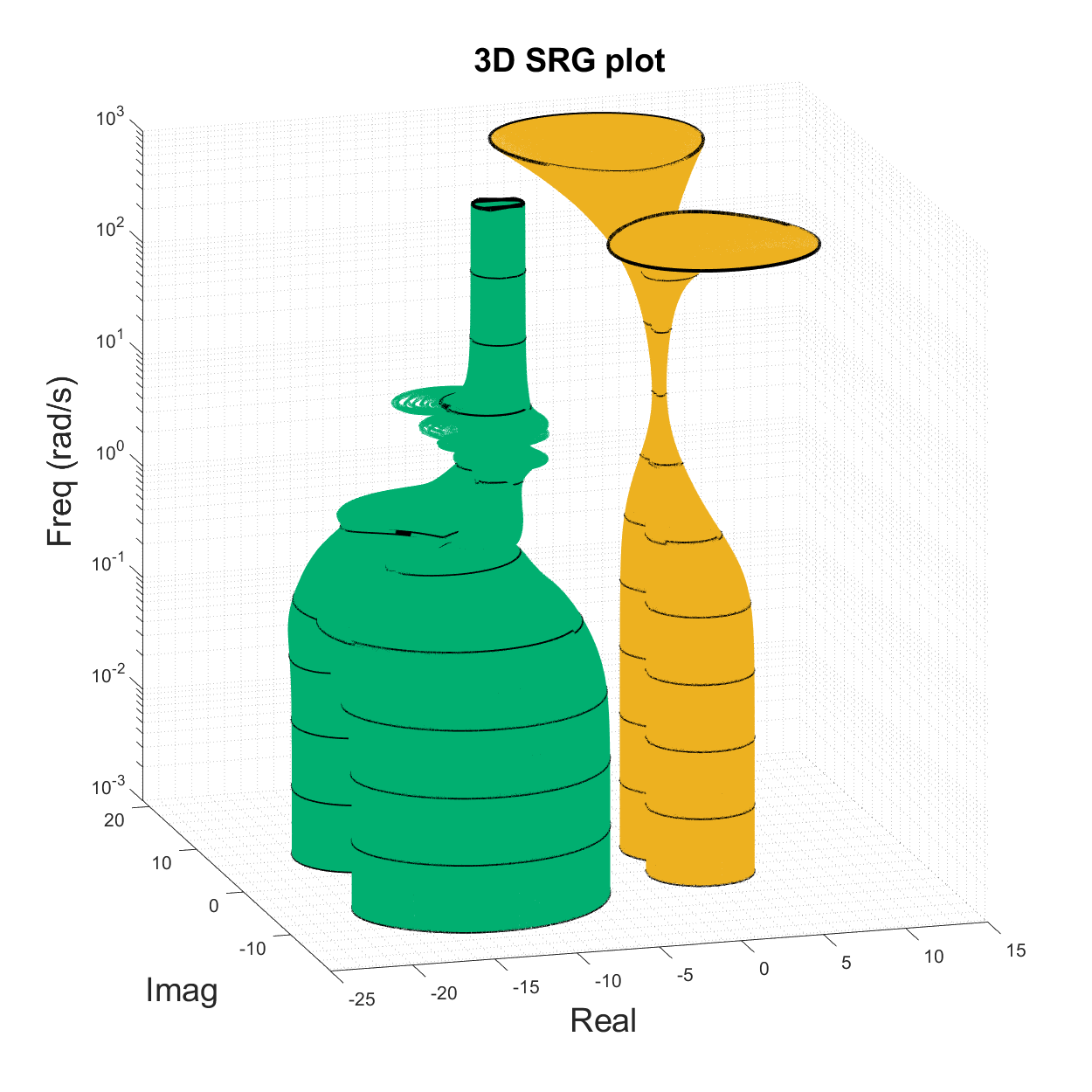}
    \caption{} \label{fig:ex2_st_3D}
\end{subfigure}  
    \caption{$\operatorname{SRG}(H(s))$  in green, $-\operatorname{SRG}(H_{3}^{-1}(s))$ in orange and $-\operatorname{SRG}(H_{4}^{-1}(s))$ in yellow $~\forall \omega\in[10^{-3},10^3]$rad/s. (a) Nyquist plot of $\det(I+H(s)H_{3}(s))$ (b) Nyquist plot of $\det(I+H(s)H_{4}(s))$   (c) 3D plot of the $\operatorname{SRG}(H(s))$ and $-\operatorname{SRG}(H_{3}^{-1}(s))$ (d) 3D plot of the $\operatorname{SRG}(H(s))$ and $-\operatorname{SRG}(H_{4}^{-1}(s))$.}
    \label{fig:SRGExamples2}
\end{figure}
 In this example, the feedback loop of $H(s)$ and $H_3(s)$ is unstable, while the feedback loop of $H(s)$ and $H_4(s)$ is stable. Figs \ref{fig:ex2_uns_nyquist} and \ref{fig:ex2_uns_3D} show the Nyquist and SRG plots of the feedback loop between $H(s)$ and $H_{3}(s)$. Furthermore, we depict the same plots for the feedback loop between $H(s)$ and $H_{4}(s)$ in Figs \ref{fig:ex2_st_nyquist} and \ref{fig:ex2_st_3D}.  Figs. \ref{fig:ex2_uns_nyquist} and \ref{fig:ex2_st_nyquist} show two Nyquist plots for MIMO systems using the GNC. It is possible to observe that the Nyquist plot does multiple loops around the origin in both plots, making it particularly difficult to use Theorem \ref{thm:GNC} to certify the stability of both closed-loop systems. When we examine Fig. \ref{fig:ex2_uns_3D}, it is possible to observe intersections between both SRGs along the entire frequency spectrum, where we can not conclude that the closed loop system with $H_{3}(s)$ is stable. 

As in Fig. \ref{fig:ex2_uns_nyquist}, the Nyquist plot in Fig. \ref{fig:ex2_st_nyquist} shows multiple loops around the origin, and thus it fails to satisfy the conditions of Theorem \ref{thm:sufficient_GNC} for ensuring stability, as the Nyquist plot exhibits crossings with the negative real axis. 

Additionally, Fig. \ref{fig:ex2_st_3D} illustrates the 3D plot of $\operatorname{SRG}(H(s))$ and $-\operatorname{SRG}(H_{3}^{-1}(s))$. When scaling $\tau\operatorname{SRG}(H_{3}^{-1}(s))$ for all $\tau \in (0,1]$, there are no intersections between the SRGs, therefore Theorem \ref{thm:GFTLTI} shows stability. 
{Furthermore, since Fig. \ref{fig:ex2_st_nyquist} confirms that the winding number around the origin is zero,} the stability of the feedback loop can still be guaranteed by Theorem \ref{thm:GNC}.  This empirical evidence suggests that SRG separation with $\tau=1$ might be sufficient for guaranteeing $\mathcal{L}_2$ stability. %Theoretical verification of SRG separation as necessary and sufficient condition for stability remains an important open question.

%The fact that the SRGs remain non-intersecting across all frequencies (Fig. \ref{fig:ex2_st_3D}) leads us to conjecture that sufficient separation between SRGs may, in fact, guarantee $\mathcal{L}_2$ stability. While preliminary, this observation suggests a deeper, potentially fundamental relationship between geometric separation in the SRG framework and input-output stability—one that merits thorough theoretical exploration.
  
%Moreover, the SRGs do not intersect at any frequency (see Fig. \ref{fig:ex2_st_3D}), suggesting that the separation between SRGs might be sufficient for guaranteeing $\mathcal{L}_2$ stability, while these observations are compelling, they remain preliminary, pointing to a promising direction for deeper investigation.

\section{Conclusions}\label{sec:conclusions}
We derive an alternative to the frequency-wise version of the GNC using SRGs, benefiting from a decoupled approach that simplifies stability analysis. This method provides deeper insights into the system's dynamic behavior across frequencies. Future work will aim to extend this approach to more general systems, including sampled-data and data-driven settings.
\bibliography{ifacconf}

@article{pates2021scaled,
  title={The scaled relative graph of a linear operator},
  author={Pates, Richard},
  journal={arXiv preprint:2106.05650},
  year={2021}
}

@article{Chaffey_2023,
title={Graphical Nonlinear System Analysis},
journal={IEEE Transactions on Automatic Control},
publisher={Institute of Electrical and Electronics Engineers (IEEE)},
author={Chaffey, Thomas and Forni, Fulvio and Sepulchre, Rodolphe},
year={2023}}

@book{ryu2022large,
title={Large-Scale Convex Optimization: Algorithms \& Analyses via Monotone Operators},
author={Ryu, E.K. and Yin, W.},
isbn={9781009191067},
year={2022},
publisher={Cambridge Univ. Press}
}

@ARTICLE{1102280,
  author={Desoer, C. and Yung-Terng Wang},
  journal={IEEE Transactions on Automatic Control}, 
  title={On the generalized {N}yquist stability criterion}, 
  year={1980}}

@article{huang2024gain,
  title={Gain and phase: Decentralized stability conditions for power electronics-dominated power systems},
  author={Huang, Linbin and Wang, Dan and Wang, Xiongfei and Xin, Huanhai and Ju, Ping and Johansson, Karl H and D{\"o}rfler, Florian},
  journal={IEEE Transactions on Power Systems},
  year={2024},
  publisher={IEEE}
}

@article{Ryu_2021,
   title={Scaled relative graphs: nonexpansive operators via 2D Euclidean geometry},
   volume={194},
   ISSN={1436-4646},
   number={1–2},
   journal={Mathematical Programming},
   publisher={Springer Science and Business Media LLC},
   author={Ryu, Ernest K. and Hannah, Robert and Yin, Wotao},
   year={2021},
   month={jun}, pages={569–619} }

@book{zhou1998,
  title={Essentials of robust control},
  author={Zhou, Kemin and Doyle, John Comstock},
  volume={104},
  year={1998},
  publisher={Prentice hall Upper Saddle River, NJ}
}

@book{millman1993,
  title={Geometry: a metric approach with models},
  author={Millman, Richard S and Parker, George D},
  year={1993},
  publisher={Springer Science \& Business Media}
}

@book{khalil2002,
  title={Nonlinear Systems},
  author={Khalil, H.K.},
  isbn={9780130673893},
  lccn={95045804},
  year={2002},
  publisher={Prentice Hall}
}

@ARTICLE{Wang2024,
  author={Wang, Dan and Chen, Wei and Qiu, Li},
  journal={IEEE/CAA Journal of Automatica Sinica}, 
  title={The First Five Years of a Phase Theory for Complex Systems and Networks}, 
  year={2024},
  volume={11},
  number={8},
  pages={1728-1743}}

@INPROCEEDINGS{Marani2022,
  author={Marani, Yasmine and Telegenov, Kuat and Feron, Eric and Kirati, Meriem-Taous Laleg},
  booktitle={2022 IEEE/AIAA 41st Digital Avionics Systems Conference}, 
  title={Drone reference tracking in a non-inertial frame: control, design and experiment}, 
  year={2022},
  volume={},
  number={}}

@book{skogestad2005,
  title={Multivariable feedback control: analysis and design},
  author={Skogestad, Sigurd and Postlethwaite, Ian},
  year={2005},
  publisher={john Wiley \& sons}
}

@article{Chen2022,
  title={A phase theory of {MIMO} {LTI} systems},
  author={Chen, Wei and Wang, Dan and Khong, Sei Zhen and Qiu, Li},
  journal={arXiv preprint:2105.03630},
  year={2021}
}

@article{chaffey2024homotopy,
  title={A homotopy theorem for incremental stability},
  author={Chaffey, Thomas and Kharitenko, Andrey and Forni, Fulvio and Sepulchre, Rodolphe},
  journal={arXiv preprint:2412.01580},
  year={2024}
}

@article{GRIGGS_2012_sufficientNyquist,
title = {On interconnections of “mixed” systems using classical stability theory},
journal = {Systems \& Control Letters},
year = {2012},
author = {Wynita M. Griggs and S. Shravan K. Sajja and Brian D.O. Anderson and Robert N. Shorten},
}

@article{Fromion_1996,
title = {A link between input-output stability and Lyapunov stability},
journal = {Systems \& Control Letters},
year = {1996},
author = {V. Fromion and S. Monaco and D. Normand-Cyrot},
}

@INPROCEEDINGS{Jonsson2001,
  author={Jonsson, U.T. and Chung-Yao Kao and Megretski, A.},
  booktitle={American Control Conference (ACC).}, 
  title={A semi-infinite optimization problem in harmonic analysis of uncertain systems}, 
  year={2001}}

@INPROCEEDINGS{Rantzer2022,
  author={Gr{ö}nqvist, Johan and Rantzer, Anders},
  booktitle={2022 European Control Conference (ECC)}, 
  title={Integral Quadratic Constraints for Neural Networks}, 
  year={2022}}

@ARTICLE{Rantzer1997,
  author={Megretski, A. and Rantzer, A.},
  journal={IEEE Transactions on Automatic Control}, 
  title={System analysis via integral quadratic constraints}, 
  year={1997}}

@article{hannah2016scaled,
  title={Scaled relative graph},
  author={Hannah, Robert and Yin, W},
  journal={UCLA CAM report},
  year={2016}
}

@article{de2025dissipativity,
  title={A Dissipativity Framework for Constructing Scaled Graphs},
  author={de Groot, Timo and Eijnden, Sebastiaan van den and others},
  journal={arXiv preprint arXiv:2507.08411},
  year={2025}
}

@ARTICLE{Baron2025_decentralized,
  author={Baron-Prada, Eder and Anta, Adolfo and Dörfler, Florian},
  journal={IEEE Control Systems Letters}, 
  title={On Decentralized Stability Conditions Using Scaled Relative Graphs}, 
  year={2025},
  volume={9},
  number={},
  pages={691-696}}

@article{chen2025graphicaldominanceanalysislinear,
      title={Graphical Dominance Analysis for Linear Systems: A Frequency-Domain Approach}, 
      author={Chao Chen and Thomas Chaffey and Rodolphe Sepulchre},
      year={2025},
      eprint={2504.14394},
       journal={arXiv preprint arXiv:2504.14394},
}

@article{scherer2000linear,
  title={Linear matrix inequalities in control},
  author={Scherer, Carsten and Weiland, Siep},
  journal={Lecture notes, dutch institute for systems and control, Delft, the Netherlands},
  volume={3},
  number={2},
  pages={62--74},
  year={2000}
}
\appendix

\section{Proof Theorem \ref{thm:GFTLTI}} \label{proof:GFTLTI}

\ebr{The Fourier basis diagonalizes any LTI operator; hence the transfer functions $H_1(s)$ and $H_2(s)$ admit a per-frequency analysis, following directly from linearity and time-invariance \cite{zhou1998,khalil2002}.}
%Since $H_1(\textup{j}\omega)$ and $H_2(\textup{j}\omega)$ are transfer functions, we can analyze their stability at each frequency by applying the superposition principle \cite{zhou1998,khalil2002}. 
Hence, we consider the following operator classes for each $\omega \in [0, \infty)$
\begin{align*}
    \mathcal{A}_{1,\omega} :=  H_1(\textup{j}\omega) ,\; 
    \mathcal{A}_{2,\omega} :=  H_2(\textup{j}\omega).\; 
\end{align*}
We apply Theorem \ref{thm:GFT} to each pair $\mathcal{A}_{1,\omega}$ and $\mathcal{A}_{2,\omega}$. Since $H_1(\textup{j}\omega)$ and $H_2(\textup{j}\omega)$ belong to $\mathcal{RH}_\infty^{m\times m}$, the conditions for applying the theorem are satisfied (finite incremental gain {\cite[Section 4.2]{zhou1998}}). Using \eqref{eqn:SRGEqui}, we obtain
\begin{align}
         \operatorname{SRG}(\mathcal{A}_{1,\omega})^{-1} \cap -\tau \operatorname{SRG}(\overline{\mathcal{A}}_{2,\omega}) = \emptyset, \label{eqn:GFTLTI_step_1}
\end{align}
for each  $ \omega \in [0, \infty)$. Since $H_1(\textup{j}\omega)$ or $H_2(\textup{j}\omega)$ has the chord property for each  $ \omega \in [0, \infty)$,  \eqref{eqn:GFTLTI_step_1} is equivalent to
\begin{align}
     \operatorname{SRG}({H}_1(\textup{j}\omega))^{-1} \cap -\tau \operatorname{SRG}({{H}}_2(\textup{j}\omega)) = \emptyset,
    \label{eqn:proofcor}
\end{align}
for each  $ \omega \in [0, \infty)$. $\hfill\blacksquare$
\section{Proof Theorem \ref{thm:equivalence}} \label{proof:equivalence}
Now we prove 1)$\Rightarrow$2).  We recall the Minkowski sum of two sets $A$ and $B$ as $A + B = \{a+b|\; a \in A,\; b \in B\}.$ In addition, we recall the frequency-wise version of the GNC as 
$$\det (I+\tau H_{1}(s)H_{2}(s))\neq 0,\; \forall \tau \in\;(0,1].$$
Since $H_1(s)$ is invertible, a multiplication of the GNC with $\det(H_{1}^{-1}(s))$ does not modify the original condition, i.e.,
\begin{align}
    \det(H_{1}^{-1}(s))\det (I_n+\tau H_{1}(s)H_{2}(s))\neq 0. \label{eqn:step1}
\end{align}

We use the determinant properties of the Schur complement to represent \eqref{eqn:step1} as 
\begin{align*}
    \det \left(\left[\begin{matrix}   H_{1}^{-1}(s)& -\tau H_{2}(s) \\ I_n& I_n \end{matrix}  \right]\right) \neq 0,
\end{align*} 
or equivalently as $ \det(I_n)\det (H_{1}^{-1}(s)+\tau H_{2}(s))\neq 0.$ Since the determinant of a matrix is the product of its eigenvalues, we have equivalently
\begin{align}
    \prod_{i=1}^n  \lambda_i(H_{1}^{-1}(s)+\tau H_{2}(s)) \neq 0 \label{eqn:det_to_srg}.
\end{align}
Equation \eqref{eqn:det_to_srg} holds if $\lambda_i(H_{1}^{-1}(s) + \tau H_{2}(s))\neq 0 \; \forall i$. The spectrum lies within the SRG, i.e., $\operatorname{SRG}(A) \supseteq \lambda(A)$ \cite[Thm 1]{pates2021scaled}.  {Additionally, $0 \in \operatorname{SRG}(A)$ if and only if any $\lambda_i(A)=0$.} Hence,~\eqref{eqn:det_to_srg} implies  
\begin{align}
   0 &\notin \operatorname{SRG} \left( H_{1}^{-1}(s)+\tau H_{2}(s)\right), \label{eqn:lemma2_step1}
\end{align}
 {Since $H_1(s)$ or $H_{2}(s)$ have the chord property \cite[Thm 6]{Ryu_2021},  it follows that}
\begin{small}
   \begin{align*}
\operatorname{SRG}\left(  H_{1}^{-1}(s)+\tau H_{2}(s)\right) \subseteq     \operatorname{SRG}\left(  H_{1}^{-1}(s)\right) +\operatorname{SRG}\left(\tau H_{2}(s)\right). 
\end{align*}  
\end{small}

Thus, \eqref{eqn:lemma2_step1} can be written as 
\begin{align}
      0 \notin \operatorname{SRG}\left(  H_{1}^{-1}(s)\right) +\operatorname{SRG}\left(\tau H_{2}(s)\right), \label{eqn:Minkowskix2}
\end{align}
which means that for each $\omega$, there should not be any intersection between $\operatorname{SRG}\left(  H_{1}^{-1}(s) \right)$ and $-\operatorname{SRG}\left(\tau H_{2}(s) \right)$. We can rewrite \eqref{eqn:Minkowskix2} as
\begin{align} 
      \operatorname{SRG}\left(  H_{1}(s) \right)^{-1} \cap -\tau \operatorname{SRG}\left(H_{2}(s)  \right)= \emptyset, \; \label{eqn:equivalence_step3}
\end{align}
which is the GFT for LTI systems in \eqref{eqn:SRGEquiLTI}. Therefore, we can conclude that $\det (I+\tau H_{1}(s)H_{2}(s))\neq 0 \;\forall \;\tau(0,1]$ is equivalent to \eqref{eqn:SRGEquiLTI} for LTI MIMO systems $\in \mathcal{RH}_\infty$ for each $\omega \in[0, \infty)$. 

The implication 2)$\Rightarrow$1) can be established by reversing the steps used in the proof of 1)$\Rightarrow$2).
$\hfill\blacksquare$

\end{document}